\documentclass[11pt,eps]{article}
\usepackage{graphicx,amssymb,amsfonts,amsmath,amssymb,amscd,amstext, mathrsfs}
\makeatletter

\@addtoreset{equation}{section}
\newcommand{\be}{\begin{eqnarray}}
\newcommand{\ee}{\end{eqnarray}}
\newcommand{\ba}{\begin{array}}
\newcommand{\ea}{\end{array}}
\newcommand{\nn}{\nonumber}

\makeatletter \@addtoreset{equation}{section} \makeatother

\begin{document}
\vspace{1cm}
\begin{center}
~\\~\\~\\
{\bf  \LARGE Generalized Gravitational Entropy from Various  Matter Fields}

\vspace{1cm}

                      Wung-Hong Huang\\
                       Department of Physics\\
                       National Cheng Kung University\\
                       Tainan, Taiwan\\
\end{center}
\vspace{1cm}
\begin{center}{\bf  \Large ABSTRACT } \end{center}
The generalized gravitational entropy proposed in recent by  Lewkowycz  and Maldacena [1] is extended to the systems of Boson fields, Fermion fields and Maxwell fields which have arbitrary frequency and mode numbers on the BTZ  spacetime.  We find the associated regular wave solution in each case and  use it  to calculate the exact gravitational entropy.  The results show that there is a threshold  frequency below which the Fermion fields  could not contribute the generalized gravitational entropy.  Also, the static and zero-mode solutions  have no entropy,  contrast to that in scalar fields.  We also find that the entropy of  the static scalar fields and non-static fermions  is an increasing function of mode numbers and, after arriving the maximum entropy, it becomes a deceasing function approaching to a constant value.  We calculate the gravitational entropy of Maxwell fields and use the duality between EM and scalar fields to explain its result. The gravitational entropy from DBI action is also studied.
\\
\\
\\
\\
\\
\begin{flushleft}
*E-mail:  whhwung@mail.ncku.edu.tw\\
\end{flushleft}
\newpage
\pagenumbering{arabic}
\tableofcontents
\section{Introduction}
According to the Gibbons and Hawking method [2]  the thermodynamics of black holes  is studied by the Euclidean  partition function with  periodic field $\phi(\tau)= \phi(\tau+\beta)$.  
\be
Z&=&\int D g_{\mu\nu}\int  D \phi~e^{-W_{gravity}(g)-W_{matter}(g,\phi)}\nn\\
&=&\int D g_{\mu\nu}~ e^{-W_{gravity}(g)}~[e^{-W_M(g)}]\nn\\
&=&e^{-[W_g+W_M]}
\ee
The general thermal entropy  is then calculated by [3] 
\be
S_{thermal}&=& -(\beta\partial_\beta-1)~\log Z(\beta)\nn\\
&=& (\beta\partial_\beta-1)(W_g)+(\beta\partial_\beta-1)(W_M)\nn\\
&=& S_g+S_M
\ee
Therefore, the back hole total thermal entropy contains two terms.  The first term is from gravity and called as the Hawking area term.  The second term is the quantum correction part which is from the quantized matter field and can be  called as the entanglement entropy [4,5,6,7].  The entanglement entropy constitutes only a (quantum) part of the thermodynamical entropy of the black hole, which is logarithmic divergent [3,8,9]. 

It shall be noticed that in above relation the metric $g_{\mu\nu}$ is a solution to the saddle-point equation 
\be
{\delta (W_g+W_M)\over \delta g_{\mu\nu} }=0
\ee
In the standard investigations the metric $g_{\mu\nu}$ is fixed and $W_M(g)$ could be calculated in many approaches [10,11].  In other word, the back-reaction effect on the spacetime is neglected.  

In recent  Lewkowycz  and Maldacena [1] proposed the generalization of the usual black hole entropy formula [2,3,4] to Euclidean solutions without a Killing vector.  The model spacetime they considered is 
\be
ds^2={du^2\over n^{-2}+u^2}+u^2d\tau^2+(n^{-2}+u^2)d \theta^2
\ee
and the associated gravitational entropy of  complex scalar field is calculated exactly through the formula
\be
S&=&-n\partial_n\Big[\log Z(n)-n \log Z(1)\Big]_{n=1}
\ee
in which $Z(n)$ is the  on-shell Euclidean  partition function. 

On other hand, the  3D  BTZ black hole geometry  is described by [12,13] 
\be
ds^2=(r^2-\mu)dt^2+ {dr^2\over r^2-\mu}+r^2 d \theta^2
\ee
The associated black hole temperature is  $T=\beta^{-1}$ with
\be
\beta={2\pi\over \sqrt \mu}
\ee
After defining $u^2=r^2-\mu$ the BTZ geometry can also be transformed to
\be
ds^2=u^2dt^2+{du^2\over u^2+m^2}+(u^2+m^2)d \theta^2
\ee
in which we replace $\mu$, which is mass of BTZ black hole, by  $m^2$ to save the expression in below.  Above geometry  is similar to that used in [1] after replacing  $m^2\rightarrow n^{-2}$, in which the associated generalized gravitational entropy is calculated from the on-shell action $Z[n]$, which tells us the  back-reaction effect of classical field  on the BTZ black hole [1]. 

 In this paper we will continue the previous investigations [1,14], which studied only Boson fields and Maxwell fields, to the  Fermion fields. Especially, the fields consider in this paper will have arbitrary frequency and mode numbers on the BTZ  spacetime.  We will find the associated regular wave solution in each case and  use it  to calculate the exact gravitational entropy. The results are used to discuss how the  generalized gravitational entropy  is dependent of  frequency and mode numbers.

  The  generalized gravitational entropy is known to correspond to the correction of area law as shown in [1] and we derive it more detail  in section 2.   In section 3 we first present the  new,  general scalar-field gravitational entropy in Eq.(3.7) which reduces to the solutions in [1,14] in special limit.  In section 4 we exactly solve the Fermion fields equation on BTZ spacetime [15,16,17] to find the most general regular solution which has arbitrary values of  mode numbers and frequency.  We use the solution  to calculate the exact gravitational entropy in Eq.(4.21) and analyze its dependence on the mode numbers.  We  see that the the static and zero-mode solutions  have no entropy.  We also found that the entropy of  the  non-static fermions  is an increasing function of mode numbers and, after it arrive a maximum entropy it becomes a deceasing function and is derived to the  asymptotic value.  

Section 5 turns to the case of Maxwell field. We  present the  gravitational entropy in Eq.(5.9) and Eq.(5.15) which reduce to the solutions in [14] in special limit.  We show that the entropy is same in Lorentz and Coulomb gauge.  We also use the duality between EM and scalar fields to explain the similarity of  gravitational entropy between them.  The problem of normalization of  Maxwell field is discussed. In section 6 we consider the  DBI action [18] on the BTZ spacetime and find the gravitational entropy in Eq.(6.8).  We explicitly check that the classical solution which gives  the  black hole entropy precisely corrects the black hole area law in this case.   Last section is devoted to a short summary.   We  present an appendix to quickly solve the fermion fields equation on BTZ spacetime by the conformal transformal transformation [19,20,21], which is used to check the simplest solution we found in section 4.  Study of quantum fermion field on BTZ black hole can be found in [3,22,23,24] for example. 
\section{Classical Solution and Area Law}
We first follow the paper [1]  to prove that the classical solution of Einstein equation  gives  the correction of horizon area in the real black hole spacetime.  While the original proof  in [1] used the coordinate  $(t,u,\phi)$ in Eq(1.4) we will in here adopt the black hole coordinate  $(t,r,\phi)$.  As the metric in the coordinate Eq(1.6) explicitly shows the black mass $\mu$ we can easily see the temperature  in our proof. \\

Consider the matter field Lagrangian   $ {\cal L}(g_{\mu \nu},\Phi,\nabla_{\mu} \Phi) $ on  Euclidean spacetime with $t=t+\beta$.  Its thermal entropy  (It is called as generalized gravitational entropy in [1])  can be calculated by
\be
 S_T&=& -\beta \partial_\beta \Big[ \log Z^{\rm matter}\Big]+ \log Z^{\rm matter}\nn\\
& = & -\beta \partial_\beta \Big[ \int dt \int d\vec x \sqrt{g} {\cal L}_{\rm matter}\Big] +\Big( \int dt \int d\vec x  \sqrt{g} {\cal L}_{\rm matter}\Big)\nn\\
& = & -\beta^2 \partial_\beta \Big[\int d\vec x  \sqrt{g} {\cal L}_{\rm matter}\Big]\nn\\
& = & -\beta^2 \Big[\int d\vec x  \Big({\partial \sqrt{g} {\cal L}_{\rm matter} \over \partial g^{\mu \nu}}\frac{\partial g^{\mu \nu}}{\partial \beta}+ {\delta \sqrt{g} {\cal L}_{\rm matter} \over \delta \Phi} \delta (\partial_\beta \Phi)\Big)\nn\nn\\
&=&-{\beta\over 16\pi G}~ \int dt~d\vec x  \sqrt{g}~G_{\mu\nu} { \partial g^{\mu\nu } \over \partial \beta }
\ee
in which we have used matte field equation ${\delta \sqrt{g} {\cal L}_{\rm matter} \over \delta \Phi}=0$ and Einstein field equation $G_{\mu \nu}=8\pi GT_{\mu \nu}$.
\\

 To proceed, we also follow [1] to prove that  above result is just the area of the horizon.  As we consider the system on fixed metric, the probe classical matter scalar field is small, say $\phi\sim~\eta$,  then the metric will get correction $O(\eta^2)$. Thus the  gravitational part of the action has not term of order $\eta^2$ and  $\partial_\beta$ derivative of the gravitational part vanishes at order $\eta^2$. Thus
\be
 \partial_\beta [ \log Z^{\rm Grav}]&=&0= \int dt~d\vec x~\sqrt{g} G_{\mu\nu} {\partial g^{\mu\nu } \over \partial \beta }  - \int dt~d\vec y~  \sqrt{g} \nabla_\mu \partial_\beta g^{\mu r}
\ee
in which $d\vec y$ is $d\vec x$ while removes the radius coordinate $dr$. Thus
\be
 \int dt~d\vec x~\sqrt{g} G_{\mu\nu} {\partial g^{\mu\nu } \over \partial \beta }&=& \int dt~d\vec y~ \sqrt{g} \nabla_\mu \partial_\beta g^{\mu r}|_{r=r_H}
\ee
The right integration gives the surface integration, i.e, the area of the horizon, ${\cal A_H}$, or more precisely the area of the horizon at order $\eta^2$. To see this property, let us consider the following black hole spacetime with horizon radius $r_H$ at which $C(r_H)=0$
\be
ds^2=-C(r)dt^2+{dr^2\over C(r)}+\cdot\cdot\cdot
\ee
in which ``$\cdot\cdot\cdot$" does not contain coordinate $dt$ and $dr$.  Above black hole has inverse temperature $\beta$ with 
\be
C'(r_H) =4\pi \beta^{-1}
\ee
and
\be
 \int dt~d\vec y~ \sqrt{g} \nabla_\mu \partial_\beta g^{\mu r}|_{r=r_H}&=& \int dt~d\vec y~  \sqrt{g}~ \partial_\beta  (4\pi\beta^{-1})\nn\\
&=&-4\pi \beta^{-1}~\int d\vec y~  \sqrt{g_{\vec y}}\nn\\
&=&-4\pi \beta^{-1}~{\cal A_H}
\ee
Comparing to the original paper [1]  our general formula shows the explicity factor $4\pi\beta^{-1}$. Therefore
\be
S_T={1\over 4G}{\cal A_H}
\ee
 These complete our proof. 
\section {Scalar Field on BTZ Spacetime}
The action of  scalar field we will consider can be described by 
\be
A^{\Phi}=\int d^3 \sqrt g~g^{ab}\partial_a\Phi^* \partial_b\Phi=\int d^3 \partial_a[\sqrt g~g^{ab}\Phi^* \partial_b\Phi]-\int d^3~ \Phi^*\partial_a[\sqrt g~g^{ab} \partial_b\Phi]
\ee
The first bracket is the surface term and will contribute to the no-shell gravitational action which is considered later.  After the variation the second bracket gives the scalar field equation.  We extend the ansatz in [1] and adopt the following general solution 
\be
\Phi(\tau,u,\theta)&=&\eta e^{i\omega \tau}~e^{in \theta} f_{\omega,n}(u)
\ee
While the original investigation is to let mode number $n=0$ and frequency $\omega=1$ we keep them as two arbitrary values in order to see how the gravitational entropy will depend on.  

The associated differential equation of $ f_{\omega,n}(r) $ becomes 
\be
u (u^2+m^2) \Big( u (u^2 + m^2)  f_{\omega,n}''(u)+(u^2 + 3 m^2)  f_{\omega,n}'(u)\Big )-(n^2 u^2 + (u^2 +m^2) \omega^2)  f_{\omega,n}(u) =0\nn\\
\ee 
There are two independent solutions and one of them is a regular solution which is finite on horizon $u=0$. After normalized it  by the condition $f_{\omega,n}(\infty)=1$ as that in  [1] we find  
\be
f_{\omega,n}(u)&=&u^{\omega\over m}(u^2+m^2)^{in\over2m}m^{-in-\omega\over m}\Gamma\Big(1+{\omega-in\over 2m}\Big)~\Gamma\Big(1+{\omega+in\over 2m}\Big)\nn\\
&&\times~_2\tilde{F}_1\Big({in+\omega\over 2m},1+{in+\omega\over 2m},1+{\omega\over m},-{u^2\over m^2}\Big)
\ee 
Note that the action, as we analyze before, has two parts.  The second bracket become zero after put the field on-shell and it remains only the first bracket which is the surface term. Thus the classical on-shell  action becomes (hereafter we let $\eta=1$)
\be
A_{\rm on-shell}^{\Phi}&=&\int d^3x\partial_a[\sqrt g~g^{ab}\Phi^* \partial_b\Phi]=2\pi\beta \Big(  \sqrt g~g^{uu}f_{\omega,n}(u) \partial_u f_{\omega,n}(u)\Big)_{u\rightarrow\infty}
\ee
After substituting the exact solution of  $f_{\omega,n}(u)$ we find that
\be
\log Z^{\Phi}(\beta)&=&A_{\rm on-shell}^{\Phi}=-\pi \beta\Big({4in\pi\over \beta} +\gamma[n^2 + \omega^2] + (n^2 + \omega^2)\Big (H\Big[{\beta(in+\omega)\over 4\pi}\Big] \nn\\
&&- 2 \ln(u)+2\ln\Big({2\pi \over \beta}\Big) + \psi^{(0)}\Big[{\beta(-in + \omega)\over 4\pi}\Big]\Big)\Big)+ {\cal O} \Big({1\over u}\Big)
\ee
in which $\gamma (x)$ is the EulerGamma function, $\psi^{(k)}(x)$ is the PolyGamma function and H is the HarmonicNumber. The terms linear in $\beta$ include divergent terms that should be subtracted [1]. However, they do not contribute to the entropy.  The associated generalized gravitational entropy then becomes 
\be
S^{\Phi}(m,n,\omega)&=&m^{-2}\pi^2\Big(4m(imn+n^2+\omega^2)
 + (n + i\omega)^2(in+\omega) \psi^{(1)}\Big[{-in + \omega\over 2m}\Big]\nn\\
&&+  (n - i\omega)^2(-in+\omega) \psi^{(1)}\Big[1+{in + \omega\over 2m}\Big]\Big)
\ee
Let us now analyze the behaves of  above generalized gravitational entropy : 
\subsection{Zero-Model Scalar Field Solution}
   For the zero mode solution, $n=0$,  above result gives
\be
 S^{\Phi}(m,0,\omega)&=&{2 \pi^2 \omega\over m^2} \Big(2 m (m + \omega) - \omega^2 \psi^{(1)}\Big[ {\omega\over2 m}\Big]\Big)
\ee
and 
\be
S^{\Phi}(1,0,1)&=&\pi^2(8-\pi^2)
\ee
which is consistent with [1] after replacing $L_x$ in there by $2\pi$. Note that the zero-model scalar field entropy is always negative and is a deceasing function of $\omega$. We plot it in figure 1.
\\
\\
\scalebox{0.8}{\hspace{3cm}\includegraphics{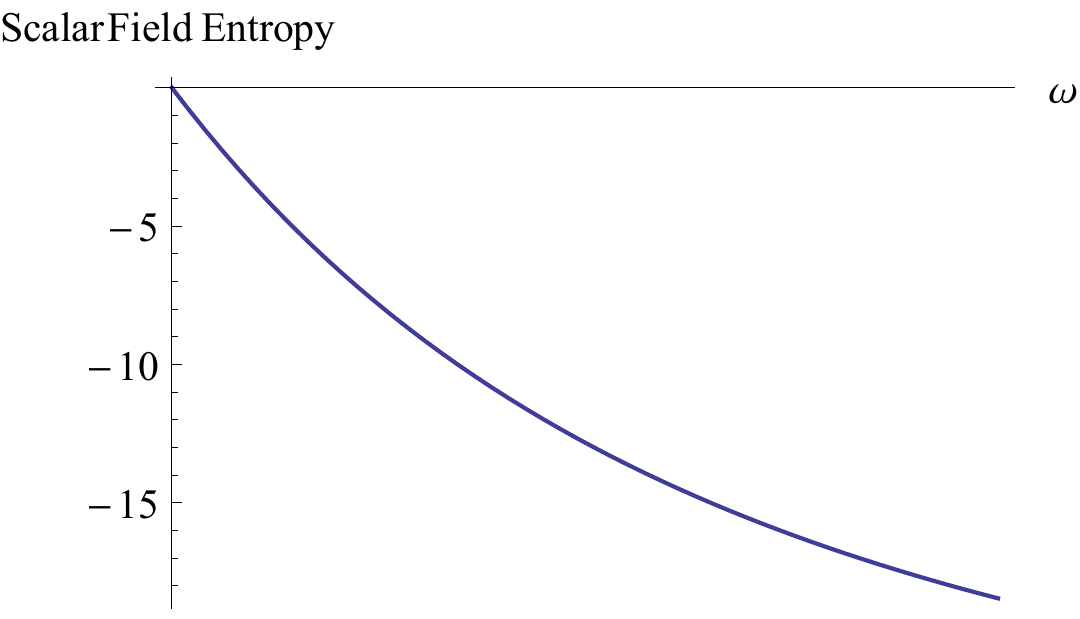}}
\\          
{\it Figure 1:  Dependence of  generalized gravitational entropy of static scalar field  solution  on the frequency  $\omega$.}
\subsection{Static Scalar Field Solution}
 For the case of static solution, $\omega=0$, we find
\be
S^{\Phi}(m,n,0)&=&m^{-2}\pi^2\Big(4m(imn+n^2)
 + in^3 \psi^{(1)}\Big[{-in \over 2m}\Big]-in^3\psi^{(1)}\Big[1+{in \over 2m}\Big]\Big)\\
&\approx&
\left\{
\ba {c}
\frac{4 \pi ^2 n^2}{m}+{\cal O} \Big(n^4\Big)\\
\frac{8 \pi ^2 m}{3}+\frac{32 \pi ^2 m^3}{15 n^2}+{\cal O} \Big(n^{-2}\Big)
\ea
\right.
\ee
which shows that the  zero mode of static solution has not generalized gravitational entropy and  the  entropy  approaches to ${8m\pi^2 \over3}$ asymptotically. The positive coefficient $\frac{32 \pi ^2 m^3}{15 n^2}$ in large $n$ expansion shows that there exists a maximum entropy in static solution. For clear, in figure 2 we plot the the  entropy of static scalar field  solution v.s. mode number  $n$. It shows that  the entropy is an increasing function of mode number $n$ for small $n$ and, after arriving a maximum entropy it becomes a deceasing function and is derived to the  asymptotic value.  
\\
\\
\scalebox{0.8}{\hspace{3cm}\includegraphics{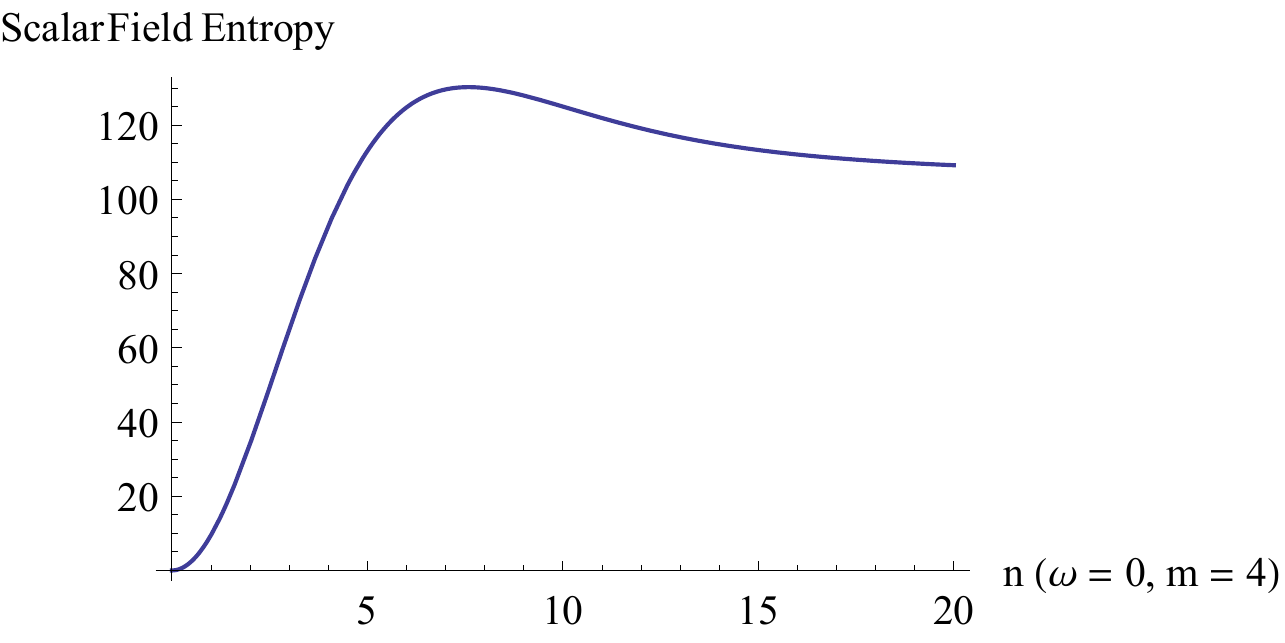}}
\\          
{\it Figure 2:   Dependence of  generalized gravitational entropy of static scalar field  solution  on the mode number $n$.}
\subsection{Non-static Scalar Field Solution}
 Consider the case of  non-static scalar field solution with $\omega \ne 0$.  Then
\be
S^{\Phi}(m,n,\omega)&\approx&
\left\{
\ba {c}
 2 \pi ^2 \omega \left( (2+{\omega\over m} )-{\omega^2\over m^2} \psi
   ^{(1)}\left(\frac{\omega }{2 m}\right)\right)+{\cal O} \Big(n^2\Big)
\\\\
{8m\pi^2 \over3}+\frac{16 \pi ^2 m \left(2 m^2-5 \omega ^2\right)}{15 n^2}+{\cal O} \Big(n^{-3}\Big)\\
\ea
\right.
\ee
Notice that  the entropy expanded in small mode number is always negative, i.e. $ (2+{\omega\over m} )-{\omega^2\over m^2} \psi^{(1)}\Big(\frac{\omega }{2 m}\Big)< 0 $, while entropy expanded  in  large  mode number is always positive, i.e. ${8m\pi^2 \over3}>0$ .  Thus, it seems that there has a special value of  $n$ which give zero entropy.  In fact, this is not the case, because that the mode number $n$ is  an integral and we cannot conclude that there exist a special mode which give zero entropy.  It is interesting to see that the  generalized gravitational entropy is negative for small $n$ and it becomes the positive value ${8m\pi^2 \over3}$ for the solution of sufficient large mode number.  

In figure 3 we plot the the  entropy of non-static scalar field  solution with $\omega=1$ and $m=1$ to explicitly see above  properties.
\\
\\
\scalebox{0.8}{\hspace{3cm}\includegraphics{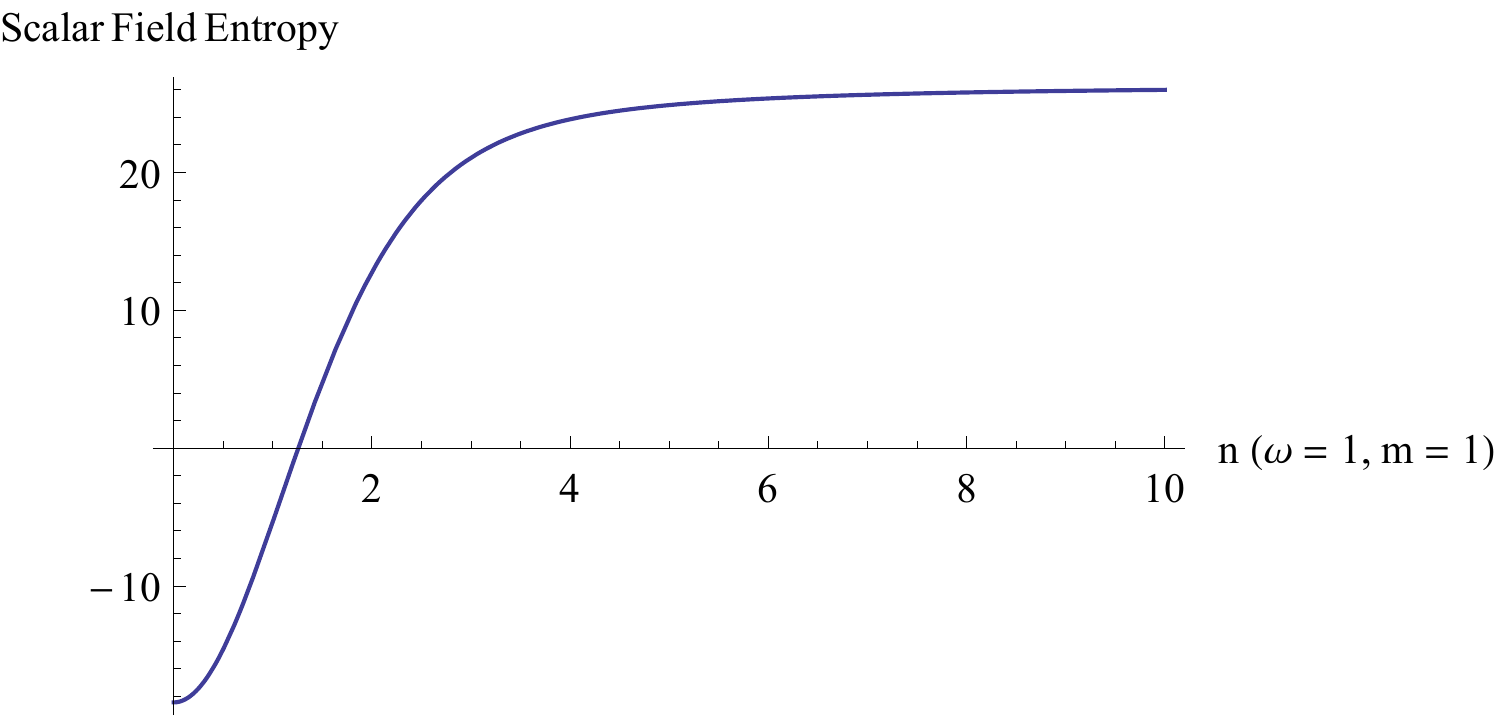}}
\\          
{\it Figure 3:   Dependence of  generalized gravitational entropy of non-static scalar field solution  on the mode number $n$.}
\section {Fermion Field on BTZ Spacetime}
The action of  Fermion  field we will consider can be described by [15,16,17]
\be
A^{\Psi}&=&-\int d^3x \sqrt g~\Big[{1\over 2}\bar\Psi \gamma^\mu (\stackrel{\rightarrow}{D_\mu}-\stackrel{\leftarrow}{D_\mu})\Psi\Big]-\int d^3x\partial_\mu[\sqrt g~\bar\Psi \gamma^\mu\Psi]
\ee
The first term is the bulk action which does not provide a full description of the dynamics. The
problem lies with the terms which arise after integrating by parts, which are evaluated on the boundary [16,17].    The second term is the corresponding surface action which will contribute to the on-shell gravitational action and is considered later.  The covariant derivative  $D_\mu$  is defined by 
\be
D_\mu&=&\partial_\mu+{1\over 4}\omega_{ab\mu}\Gamma^{ab}
\ee
To find its value we first note that the tetrad formulism on the background (1.5) is described by
\be
e^0=u d\tau,~~~~e^1=\sqrt{u^2+m^2}~d\theta,~~~e^2={1\over \sqrt{u^2+m^2}}~du
\ee
After choosing \footnote {We are considering  the Euclidean  spacetime and $\{\tilde\gamma_a,\tilde\gamma_b\}=2\eta^{ab}=2(1,1,1)$.}
\be
\tilde \gamma_0=\sigma_z,~~\tilde \gamma_1=\sigma_x,~~\tilde \gamma_2=\sigma_y
\ee
the matrix $\Gamma_{ab}={1\over 2}\{\tilde \gamma_a,\tilde \gamma_a\}$ is
\be
\Gamma_{01}=i\sigma_y,~~~\Gamma_{02}=-i\sigma_x,\Gamma_{12}=i\sigma_z
\ee
and  the generalized Dirac matrices  define by $\gamma_\mu=e^a_\mu \tilde\gamma_a$ become
\be
\gamma_\tau=u\sigma_z,~~\gamma_u={1\over \sqrt{u^2+m^2}}~\sigma_x,~~\gamma_\theta=\sqrt{u^2+m^2} ~\sigma_x
\ee
The spin connections defined by $d e^a=-\omega^a_{~b}~e^b$  are
\be
\omega^0_{~2}={ \sqrt{u^2+m^2}\over u}~e^0,~~~\omega^1_{~2}={-u\over (u^2+m^2)^{3/2}}~e^1,
\ee
and finally
\be
\gamma^\mu\omega_{ab\mu}\Gamma^{ab}=2\Big[{-u\over (u^2+m^2)^{3/2}}+{ \sqrt{u^2+m^2}\over u}\Big]~\sigma_2
\ee
Fermion on 3D background  can  be expressed in  two component spinors 
\be
\Psi=\left\{
\ba{c}
\psi_1(\tau,r,\theta)=e^{i\omega \tau+in\theta}f(u)\\
\psi_2(\tau,r,\theta)=e^{i\omega \tau+in\theta}g(u)\\
\ea
\right.
\ee
After substituting it into the field equation we find the  following two differential equations
\be
0&=&\Big(\sqrt {u^2+m^2}~\partial_u-{u\over 2(u^2+m^2)^{3/2}}+{\sqrt {u^2+m^2}\over 2u}-{n\over\sqrt {u^2+m^2}} \Big)g(u)+{\omega\over u} f(u)\\
0&=&\Big(\sqrt {u^2+m^2}~\partial_u-{u\over 2(u^2+m^2)^{3/2}}+{\sqrt {u^2+m^2}\over 2u}+{n\over\sqrt {u^2+m^2}} \Big)f(u)-{\omega\over u} g(u)
\ee
\subsection{Zero-Mode Fermion Solution}
We first consider the zero-mode solution of $n=0$.  The  solution  is
\be
f[u]=g[u]=\frac{\left(\frac{\sqrt{m^2+u^2}+m}{u}\right)^{\pm\frac{\omega}{m}}} {\sqrt[4]{u^2 \left(m^2+u^2\right)}}
\ee
(We will in appendix use  the conformal method to quickly find above solution.) The on-shell action  calculated from above solution (positive sign is regular on horizon) is
\be
A_{on-shell}^{\Psi}&=&\int d^3x\partial_\mu[\sqrt g~g^{ab}\bar\Psi \gamma^\mu\Psi]=2\pi\beta \Big( u\sqrt{u^2+m^2}f(u)g(u)\Big)_{u\rightarrow\infty}=2\pi \beta
\ee
which is near in $\beta$ and does not contribute any entropy, contrasts to that in scalar field analyzed in previous section. 
\subsection{Static Fermion Solution}
We next consider the static solution of $\omega=0$.  In this case above two equations decouple and the solutions are
\be
f(u)&=&e^{-{u\over m} tan^{-1}({u\over m})}~{1\over \sqrt u~ (m^2 + u^2)^{1/4}}\approx~{1\over m^{1/2} \sqrt u} - {m^{3/2} n \sqrt u\over m^4}\\
g(u)&=&e^{{u\over m} tan^{-1}({u\over m})}~{1\over \sqrt u~ (m^2 + u^2)^{1/4}}\approx~{1\over m^{1/2} \sqrt u} + {m^{3/2} n \sqrt u\over m^4}
\ee
which explicitly show that the static becomes singular on horizon $u=0$ and could not give finite gravitation entropy, contrasts to that in scalar field analyzed in previous section..  
\subsection{Non-static Fermion Solution}
Let us turn to find the the non-static solutions.  The results  are  
\be
f(u)&=&\frac{(u-i m)^{\frac{m-2 i n}{4 m}} u^{\frac{\omega   }{m}-\frac{1}{2}} (u+i m)^{\frac{m+2 i n-4 \omega }{4 m}} \,    _2F_1\left(\frac{\omega }{m},\frac{m-2 i n+2 \omega }{2   m};\frac{2 \omega }{m}+1;\frac{2 u}{i
   m+u}\right)}{\sqrt{m^2+u^2} \, _2F_1\left(\frac{\omega   }{m},\frac{m-2 i n+2 \omega }{2 m};\frac{2 \omega
   }{m}+1;2\right)}\\
g(u)&=&{(u-i m)^{\frac{m-2 i n}{4 m}} u^{\frac{\omega }{m}-\frac{1}{2}} (u+i m)^{-\frac{7 m-2 i n+4\omega }{4 m}}  \over m \, _2\tilde{F}_1\left(\frac{\omega }{m},\frac{m-2 i n+2 \omega }{2 m};\frac{2 \omega    }{m}+1;2\right)}\nn\\
&&\times\Big[u (i m+2 n+2 i \omega ) \, _2\tilde{F}_1\left(1+\frac{\omega }{m},\frac{\omega -i
   n}{m}+\frac{3}{2};2+\frac{\omega}{m};\frac{2 u}{i m+u}\right)\\
&&-m (m-i u) \,_2\tilde{F}_1\left(\frac{\omega }{m},\frac{m-2 i n+2 \omega }{2 m};\frac{2 \omega
   }{m}+1;\frac{2 u}{i m+u}\right)\Big]
\ee
where $ _2{F}_1$ is the hypergeometric function and  $ _2\tilde{F}_1$ is the regularized hypergeometric function.   Above solution is finite on horizon and has been proper normalized. 

The on-shell action  calculated from above solution is
\be
A_{on-shell}^{\Psi}&=&\int d^3x\partial_\mu[\sqrt g~g^{ab}\bar\Psi \gamma^\mu\Psi]=2\pi\beta \Big( u\sqrt{u^2+m^2}f(u)g(u)\Big)_{u\rightarrow\infty}\\
&=&2\pi\beta\Big[\frac{(i m+2 n+2 i \omega ) \, _2\tilde{F}_1\left(\frac{m+\omega    }{m},\frac{\omega -i n}{m}+\frac{3}{2};\frac{2 (m+\omega)}{m};2\right)}{m \, _2\tilde{F}_1\left(\frac{\omega    }{m},\frac{m-2 i n+2 \omega }{2 m};\frac{2 \omega    }{m}+1;2\right)}+i\Big]
\ee
The associated generalized gravitational entropy is 
\be
&&S^{\Psi}(m,n,\omega)=\frac{4 \pi ^2  ~_2\tilde{F}_1(B) (n+i \omega )}{m^2  ~_2\tilde{F}_1(A)}
\nn\\
&&+\frac{2 \pi ^2 (i m+2 n+2 i \omega ) }{m^3  ~_2\tilde{F}_1(A)}\left((\omega -i n)  ~_2\tilde{F}_1^{(0,1,0,0)}(B)+2 \omega 
    ~_2\tilde{F}_1^{(0,0,1,0)}(B)+\omega   ~_2\tilde{F}_1^{(1,0,0,0)}(B)\right)\nn\\
&&-\frac{2 i \pi ^2  ~_2\tilde{F}_1(B) (m-2 i n+2 \omega )}{m^3  ~_2\tilde{F}_1(A)^2} \left((\omega -i n)  ~_2\tilde{F}_1^{(0,1,0,0)}(A)+2 \omega~_2\tilde{F}_1^{(0,0,1,0)}(A)+\omega   ~_2\tilde{F}_1^{(1,0,0,0)}(A)\right)\nn\\
\ee
in which $~_2\tilde{F}_1(A)=~_2\tilde{F}_1\left(\frac{\omega}{m},\frac{m-2 i n+2 \omega }{2 m};\frac{2 \omega}{m}+1;2\right)$ and $~_2\tilde{F}_1(B)=~_2\tilde{F}_1\left(\frac{m+\omega    }{m},\frac{\omega -i n}{m}+\frac{3}{2};\frac{2 (m+\omega)}{m};2\right)$.

In figure 4 we plot the the  entropy of non-static Fermion solution v.s. mode number  $n$.  It shows that  the entropy is a increasing function of mode number $n$ and, after arriving a maximum entropy it becomes a deceasing function and is derived to the  asymptotic value, likes as the static scalar field solution analyzed in previous section.
\\
\\
\scalebox{0.8}{\hspace{3cm}\includegraphics{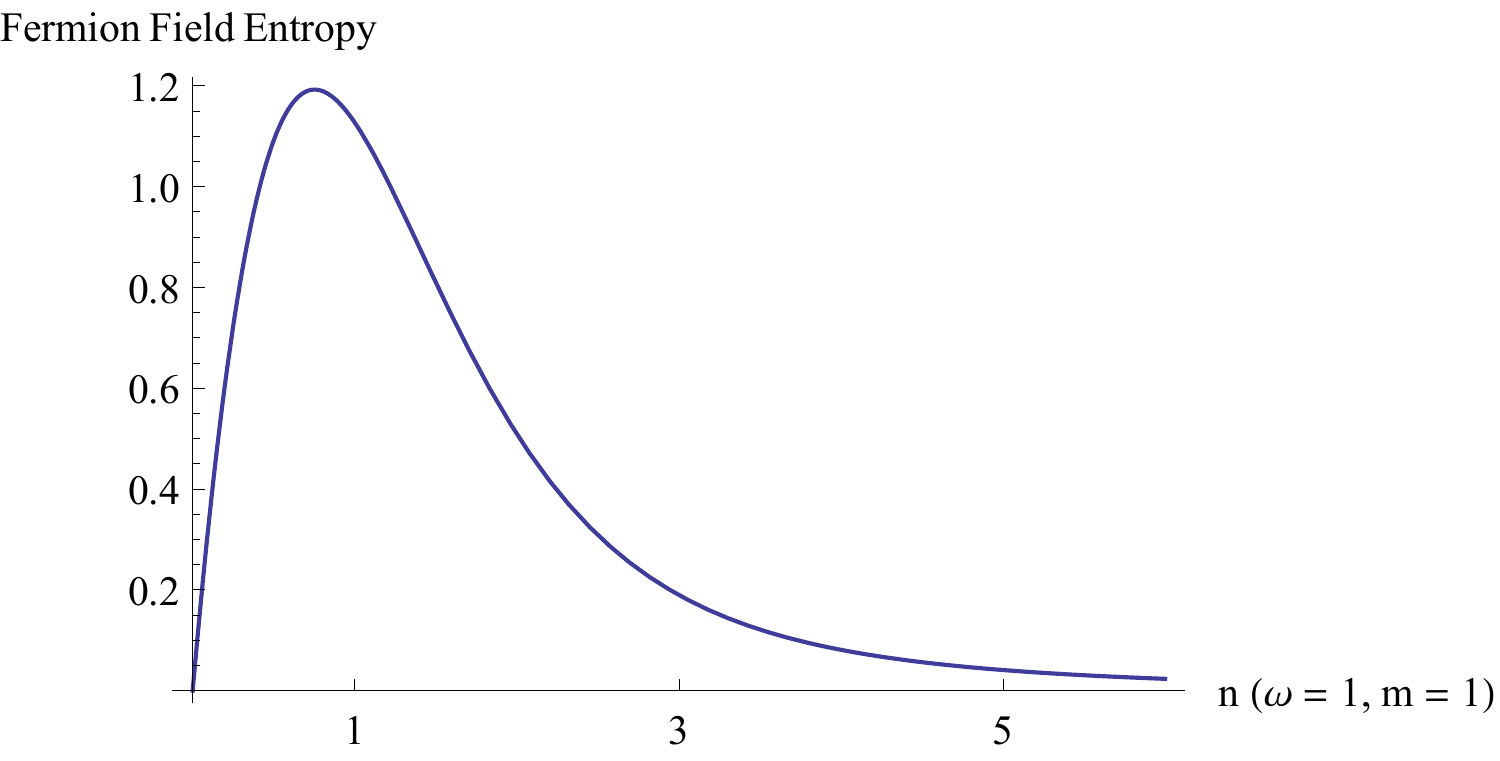}}
\\          
{\it Figure 4:   Dependence of  generalized gravitational entropy of non-static Fermion solution  on the mode number $n$.}
\\
\\
The entropy  of zero mode solution with $n=0$ can also be evaluated  from above equation and its value is zero.  Note that  near the horizon the non-static solution becomes
\be
f(u)\approx u^{{-1\over 2}+{\omega\over m}}+{\cal O} \Big(u\Big)\\
g(u)\approx u^{{-1\over 2}+{\omega\over m}}+{\cal O} \Big(u\Big)
\ee
which explicitly shows that the  solutions of  ${\omega\over m}\le{1\over 2}$ including the static case of $\omega=0$ is divergent on horizon and could not contribute the  entropy.  The result consists with the simple exact solution. 
\section {Maxwell Field on BTZ Spacetime}
In this section we  investigate the Maxwell  field solution of $A_\phi$ and $A_t$ respectively  and discuss the relations of the associated entropy with scalar field entropy from their duality.
\subsection {Maxwell Field  $A_\phi$ Solution}
The conventional action of Maxwell field is
\be
A=-{1\over 4}\int d^3x\sqrt g~g^{\mu\lambda}g^{\nu\delta}~F_{\mu\nu}F_{\lambda\delta}
\ee
We choose the Coulomb gauge of  $A_t=0$ and  assume that the Maxwell field  is 
\be
A_\mu = (0,\cos(\omega t) \cos(n \theta)A_r(u),\cos(\omega t) \cos(n \theta) A_\phi(u))
\ee
Then, the field equation of $A_r(r)$ gives trivial solution.  Therefore the action we consider becomes
\be
A&=&-{\pi\beta\over 8}\int du\Big[u A_\phi'(u)^2+{\omega ^2 A_\phi(u)^2\over u(m^2+u^2)}\Big]\nn\\
&=&-{\pi\beta\over 8}\int du~\Big[u A_\phi(u)A'_\phi(u)\Big]'
+{\pi\beta\over 8}\int du~\Big[A_\phi(u)(u A_\phi(u))'-{\omega^2 A_\phi(u)^2\over u(m^2+u^2)}\Big]\nn\\
\ee
which in indepenent of mode number.  The first bracket is the surface term and will contribute to the on-shell gravitational action, which is considered later.  After the variation the second bracket gives the  field equation 
\be
uA''_\phi(u)+A'_\phi(u)-{\omega^2 A_\phi(u)\over m^2 u + u^3}=0
\ee
The field equation has following two solutions 
\be
A^{(1)}_\phi(u)&=&u^{\frac{\omega}{m}} \, _2F_1\left(\frac{\omega}{2
 m},\frac{\omega }{2 m};1+\frac{\omega}{m};-\frac{u^2}{m^2}\right)\\
A^{(2)}_\phi(u)&=&u^{-\frac{\omega}{m}} \,
   _2F_1\left(-\frac{\omega}{2 m},-\frac{\omega}{2
   m};1-\frac{\omega}{m};-\frac{u^2}{m^2}\right)
\ee
While the second solution which being divergent on horizon is unphysical we adopt the first solution which is zero on horizon.  After normalizing the first solution by $A^{(1)}_\phi(\infty)=\log(u)$, like as in [1], we use the following function to calculate its correction to the BTZ entropy.
\be
A_\phi(u)&=&\frac{\sqrt{\pi } 2^{-\frac{\omega}{m}-1} m^{-\frac{\omega }{m}}u^{\frac{\omega}{m}} \Gamma \left(\frac{\omega }{2 m}\right) \,
   _2F_1\left(\frac{\omega}{2 m},\frac{\omega }{2 m};\frac{\omega
   }{m}+1;-\frac{u^2}{m^2}\right)}{\Gamma \left(\frac{m+\omega}{2
   m}\right)}
\ee
Substituting the solution into the on-shell gravitational action we find 
\be
\log Z(\beta)&=&-{\pi\beta\over 2\omega}~\Big(\omega  \log \left(m^2\right)+2 (m-\omega \log (u))+2\omega  \left(\psi ^{(0)}\left(\frac{\omega }{2 m}\right)+\gamma\right)\Big)
\ee
It gives  gravitational entropy 
\be
S_{A}(m,\omega,\beta)&=&\frac{\pi ^2 \left(2 m (m+\omega)-\omega^2 \psi^{(1)}\left(\frac{\omega}{2 m}\right)\right)}{m^2\omega}
\ee 
After multiple  the frequency $\omega^2 $ above result exactly matches with the half value entropy of complex scalar field in eq.(3.8)   This is because that on three dimension the Maxwell field is dual to the real scalar field, as mentioned in our previous paper [14] in which we studied only the case of $m=\omega=1$.  The reason of why the Maxwell field entropy is not exactly equal to the real scalar field entropy while it needs an extra factor of $\omega^2$ is because that the normalization constants used between them are different. See the more discussions in following section.
\subsection {Maxwell Field  $A_t$ Solution}
Let us turn to another possible solution and assume that the Maxwell field  is 
\be
A_\mu = (\cos(\omega t) \cos(n \theta)A_t(r),\cos(\omega t) \cos(n \theta) A_r(r),0)
\ee
in which we us the coordinate r not u in order to find that exact solution. The field equation of $A_r(r)$ gives trivial solution and  the action we will consider becomes
\be
A&=&-{\pi\beta\over 8}\int dr\Big[r A_t'(r)^2+{n^2 A_t(r)^2\over r(-m^2+r^2)}\Big]\nn\\
&=&-{\pi\beta\over 8}\int dr~\Big[r A_t(r)A'_t(r)\Big]'
+{\pi\beta\over 8} \int dr~\Big[A_t(r)(r A_t(r))'-{n^2 A_t(r)^2\over r(-m^2+r^2)}\Big]\nn\\
\ee
which is indepenent of frequency.  The first bracket is the surface term and will contribute to the on-shell gravitational action, which is considered later.  After the variation the second bracket gives the  field equation 
\be
uA''_t(r)+A'_t(r)-{n^2 A_t(r)\over -m^2 r + r^3}=0
\ee
The field equation has following two solutions 
\be
A^{(1)}_\phi(u)&=&r^{\frac{-i n}{m}} \, _2F_1\left(\frac{-in}{2
 m},\frac{-in }{2 m};1-\frac{in}{m};\frac{r^2}{m^2}\right)\\
A^{(2)}_\phi(u)&=&r^{\frac{i n}{m}} \, _2F_1\left(\frac{in}{2
 m},\frac{in }{2 m};1+\frac{in}{m};\frac{r^2}{m^2}\right)
\ee
Both  solutions are finite on horizon and we have to make a linear combination of them to get a function which  become zero on horizon.  After normalizing  the function  by $A_t(\infty)=\log(t)$, like as in [1], and  substituting it into the on-shell gravitational action we find that it gives the following gravitational entropy 
\be
S_{A}(m,n,\beta)&=&(n m)^{-2}\pi^2\Big(4m(imn+n^2)
 + in^3 \psi^{(1)}\Big[{-in \over 2m}\Big]-in^3\psi^{(1)}\Big[1+{in \over 2m}\Big]\Big)\nn\\
&\approx& {2\pi^2\over m}+{\cal O}(n^2)
\ee
Above result exactly matches with the half value entropy of complex scalar field in eq.(3.10) after multiple the mode number $n^2$.  This reflects the fact that  on three dimension the Maxwell field is dual to the real scalar field, as mentioned  before and discussed in following section. 
\subsection {Duality of Maxwell Field and Normalization}
Previous sections have found an interesting property that, contrasts to the scalar field and Fermion field the static solution of zero-mode Maxwell field $A_t$ can give finite entropy ${2\pi^2\over m}$. We also find that the Maxwell entropy is just the half value entropy of complex scalar field in eq.(3.10) after multiple the mode number $n^2$ or frequency $\omega$.  The property of  half value entropy is the result of the duality between Maxwell field and real scalar field  
\be
 F^{\mu\nu} =\epsilon^{\mu\nu\lambda}\partial_\lambda \Phi
\ee 
As mentioned before, the extra factor of  mode number $n^2$ or frequency $\omega$ in Maxwell field is the consequence of the normalization.  Then, there arises a problem : does the normalization is natural ? To answer this problem we can study the simplest solution of zero-mode static Maxwell solution. 

 Let us first  try the possible static zero-mode solution
\be
A_\mu = (0,0,A_\phi(u))
\ee
It has general solution
\be
A_\phi(u)=C_1+C_2\log(u)
\ee
The properties of regularity and being zero on horizon lead to  $A_\phi=0$.  Thus it could not contribute finite entropy. 

 So, let us turn to another possible static zero-mode solution 
\be
A_\mu = (A_t(u),0,0)
\ee
The field equation becomes
\be
uA_t''(u)+A_t'(u)=0
\ee
and normalized solution is 
\be
A_t(u)={1\over 2} \log(u^2+m^2)-{1\over 2} \log(m^2)
\ee
Substituting the solution into the on-shell gravitational action we find 
\be
\log Z(\beta)&=& {\pi \beta\over 2} \log(u^2+m^2)-{\pi \beta\over 4} \log(m^2)
\ee
Therefore the black hole entropy  is 
\be
S_{A}(m,0,\beta)&=&{2 \pi^2 \over m}
\ee 
which matches with eq.(5.15).  Above matching is special because that in the static zero-mode solution there is no factor of mode number $n$  nor frequency $\omega$ which we can use to normalize the solution. In next section we will  see that above simple entropy also shows in DBI action.
\subsection {Maxwell Field in Lorentz Gauge}
In this section we will study the Maxwell field  in the Lorentz gauge and compare it with that on Coulomb gauge.  To proceed we consider the action 
\be
A&=&- {1\over 4}\int d^3x {\sqrt g}~\Big[(\nabla_\mu A_\nu-\nabla_\nu A_\mu)(\nabla^\mu A^\nu-\nabla^\nu A^\mu)\Big]\nn\\
&=&- {1\over 2}\int d^3x {\sqrt g}~\Big( \nabla_\mu \Big[A_\nu(\nabla^\mu A^\nu-\nabla^\nu A^\mu)\Big] + {1\over 2}\int d^3x {\sqrt g}~\Big(A^\nu \nabla_\mu \Big[\nabla^\mu A^\nu-\nabla^\nu A^\mu\Big]\Big)\nn\\
&=&- {1\over 2}\int d^3x {\sqrt g}~\Big( \nabla_\mu \Big[A_\nu(\nabla^\mu A^\nu-\nabla^\nu A^\mu)\Big] + {1\over 2}\int d^3x {\sqrt g}~A_\mu \Big(\nabla^2~\delta^\mu_\nu- R^{\mu}_{\nu}-\nabla^\mu \nabla_\nu\Big)A^\nu\nn\\
\ee
in which $\nabla_\mu A_\nu \equiv \partial_\mu A_\nu -\Gamma_{\mu\nu}^\lambda A_\lambda$ and $[D_\mu, D_\nu] A_\lambda=R_{\mu\nu\lambda}^c A_c$.  The first bracket is the surface term and will contribute to the on-shell gravitational action, which is considered later.  After the variation the second bracket gives the  field equation.  

Therefore, in the Lorentz gauge, $\nabla_\nu A^\nu=0$, the field equation becomes
\be
(\nabla^2~\delta^\mu_\nu- R^{\mu}_{\nu})A^\nu=0
\ee
in which $R^{\mu}_{\nu}={\rm Diagonal}(-2,-2,-2)$.  

To proceed, we can choose the Maxwell as before and see that it automatically satisfies the Lorentz gauge.   In this case the associated field equation  is exactly that in Coulomb gauge. Substituting  the solution into the above on-shell gravitational action we get the same entropy as that in Coulomb gauge.  We thus conclude that the generalized black hole entropy calculated in Coulomb gauge is same as that calculated in Lorentz gauge.
\section {DBI on BTZ Spacetime }
\subsection {Classical Solution of DBI on BTZ spacetime and Entropy}
We now study the back-reaction effect on BTZ black hole by D branes, which is described by DBI action [18].  In order to find the exact solution we will consider the case with simplest  EM field 
\be 
A_\mu =(A_t(u),0,0)
\ee
Then, DBI action becomes  
\be
A&=&\int d^3x~\sqrt g~\Big[{1\over \pi^2\alpha'^2} \Big(\sqrt{g_{\mu\nu} +2\pi\alpha' F_{\mu\nu}}-\sqrt {g_{\mu\nu}} \Big)\Big]\\
&=&{1\over \pi^2\alpha'^2}~\pi\beta \int_0^\infty du\Big(-u + \sqrt{u^2 + 4\pi^2 \alpha'^2 (m^2 + u^2)  A_t'(u)^2}\Big)
\ee
and the associated conjugate momentum $\Pi$ is an integration constant which has a simple relation
\be
\Pi={1\over \pi^2\alpha'^2}{4\pi^2 (m^2 + u^2) \alpha'^2 A_t'(u)\over{\sqrt{u^2 + 4\pi^2\alpha'^2  (m^2 + u^2) A_t'(u)^2}}}
\ee
It gives solution
\be
A_t(u)=c_1+\frac{\Pi \log \left(4\alpha'\left(\sqrt{\pi ^2 \alpha'^2
   \left(m^2+u^2\right)-\Pi^2}+2\pi  \alpha' \sqrt{m^2+u^2}\right)\right)}{4\pi ^2 \alpha'^2}
\ee
where $c_1$ is another  integration constant. As before, by requiring that the solution $A_\phi(u)$ is zero on horizon and becomes $A_t(\infty)=\log (u)$  it is found that
\be
\Pi=4\pi^2 \alpha^2
\ee
Substituting it into the Lagrangian the on-shell gravitation action  is
\be
\log Z(\beta)_{DBI}&=&{1\over \pi^2\alpha'^2}\Big[{m\over 4}\Big(m - \sqrt{m^2 -4 \pi^2 \alpha'^2}\Big)-\pi^2 \alpha'^2   \log\Big({m + \sqrt{m^2 -4 \pi^2 \alpha'^2}\over 2u}\Big)\Big]
\ee
While there are divergent terms  ($r\rightarrow \infty$), which should be subtracted [1], they will do not contribute to the entropy and the associated black hole entropy  becomes 
\be
S_{DBI}(\beta)&=&-\partial_\beta\log Z(\beta)_{DBI}+\log Z(\beta)_{DBI}\nn\\
&=&{2m\over\alpha'^2} \left(-2 + 2 m\sqrt{1\over m^2 -4 \pi^2 \alpha'^2}~\right)\nn\\
&\approx&\frac{\pi^2}{m}+\frac{3\pi^4 \alpha'^2}{ m^3}+\frac{10 \pi^6 \alpha'^4}{m^5}+{\cal O}(\alpha'^6)
\ee
We thus find that the leading term is just the zero-mode entropy, (eq.5.23), found in previous section of Maxwell field theory.  This is consistent with  the property that the leading order of DBI action is just the Maxwell field.  Note that the correction entropy form DBI is positive and is a decreasing function with respective to $m$, like as that in Maxwell case.  However, if $m \rightarrow \pi \alpha'$  then the correction  is divergent, as show in figure 5. The divergence shall not be shown in physics because that the back-reaction about the BTZ space is small in prior assumption.
\\
\\
\scalebox{0.8}{\hspace{5cm}\includegraphics{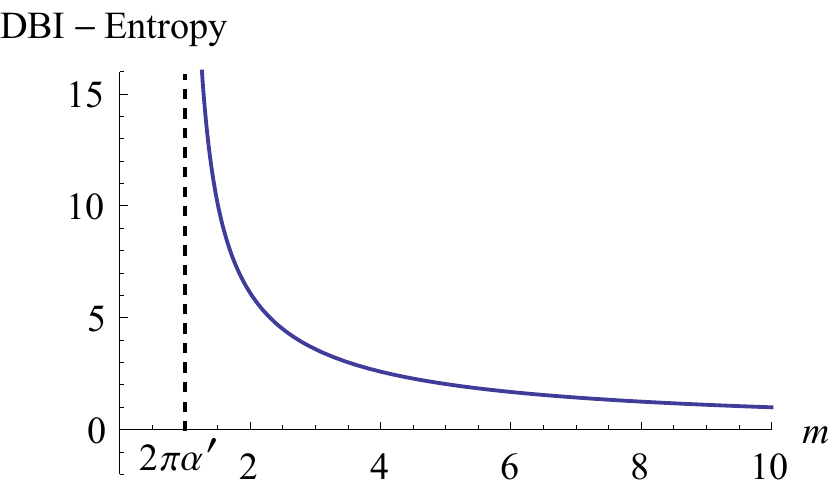}}
\\ 
{\it Figure 5:  Dependence of DBI entropy on $m$ in unit of  $2\pi \alpha'=1$.}
\subsection {Classical Solution of DBI on BTZ spacetime and Area Law}
In this section we follow [1] to explicitly check that the classical solution which gives  the  black hole entropy precisely corrects the black hole area law.  While the original paper [1] checked the case of scalar field theory and our previous paper [14]  checked the case of  Maxwell field, we will now  check  the backreaction of the DBI field on the BTZ metric.  

The action is 
\be
-A=\int d^3x~\sqrt g~\Big[R-\Lambda -{1\over \pi^2\alpha'^2} \Big(\sqrt{g_{\mu\nu} +2\pi\alpha' F_{\mu\nu}}-\sqrt {g_{\mu\nu}} \Big)\Big]
\ee
with $\Lambda=2$. The Einstein equation  is
\be
G_{\mu\nu}\equiv R_{\mu\nu}-{g_{\mu\nu}\over 2}\Big(R-2\Big)=8\pi G~T_{\mu\nu}
\ee
Consider the case  in coordinate $(t,u,\phi)$.  After the ansatz that $A_\mu=(A_t(u),0,0)$ the stress tensors become
\be
T^u_u&=&{u^2\over 2\sqrt{u^2+4 \pi^2\alpha'^2A'_t(u)^2(u^2+m^2)}}\\
T^\phi_\phi&=&{1\over2}\sqrt{u^2+4 \pi^2\alpha'^2A'_t(u)^2(u^2+m^2)}
\ee
To proceed we see that the backreaction of DBI field will  modify the  BTZ metric, which can be written as [1]
\be
ds^2 =u^2dt^2+{du^2\over u^2+m^2+g(u)}+(u^2+m^2)(1+v(u))d\phi^2
\ee 
To the leading of perturbation (small $g(u),v(u)$)
\be
G_u^u&=&{g'(u)\over  2u}\\
G_\phi^\phi&=&{g(u)\over u^2+m^2}+{(u^2+m^2)v'(u)\over 2u}
\ee
Now, using the Einstein equations we can find that 
\be
v'(u)&=&\Big({g(u)\over u^2+m^2}\Big)'+8\pi G ~{u~\partial_uA'_t(u))^2\over \sqrt{u^2+4 \pi^2\alpha'^2A'_t(u)^2(u^2+m^2)}}
\ee
To proceed we can use following  two properties to perform the integration to find  the function $v(u)$ : 

(i) Using the property that $g(0) = 0$ due to the regularity condition for the metric at the origin and $g(r)/r^2\rightarrow 0$ at infinity we can get ride of  $g(u)$ (which is a total derivative) in integration [1].

(ii) We had derived in section 6.1 that 
\be
\Pi&=&4\pi^2 \alpha'^2=\frac{4 \pi ^2 \alpha'^2 \left(m^2+u^2\right)At'(u)}{\sqrt{u^2+4 \pi ^2 \alpha'^2
   \left(m^2+u^2\right)At'(u)^2}}\\
A_t'(u)&=&\frac{2\pi \alpha'^2 u}{\sqrt{\alpha'^4 \left(m^2+u^2\right) \left(u^2-4 \pi ^2 \alpha'^2+m^2\right)}}
\ee
Therefore
\be
v(0)&=&4\pi G~\int_{0}^\infty du~ \frac{4 \pi ^2 \alpha ^2 u^2}{\left(m^2+u^2\right) \sqrt{ \left(m^2+u^2\right) \left(\left(m^2+u^2\right)-4 \pi^2 \alpha^2\right)}}\\
&=&4\pi G~\frac{-4 \pi ^2 \alpha ^2 K\left(1-\frac{4 \pi ^2 \alpha ^2}{m^2}\right)+m^2 E\left(1-\frac{4 \pi ^2\alpha ^2}{m^2}\right)-m \sqrt{m^2-4 \pi ^2 \alpha ^2} E\left(\frac{m^2}{m^2-4 \pi ^2 \alpha
   ^2}\right)}{\alpha ^2 m}\nn\\
\ee
where $E$ and $K$ are the Elliptic functions. Using the black hole area law formula 
\be
S_{DBI}&=&{1\over 4G} {\cal A_H}={1\over 4G} \cdot 2\pi \cdot m\cdot  \Big(1+v(0)\Big)=S^{(0)}_H+\delta S_H
\ee
we finally find that 
\be
\delta S_H ={1\over 4G} \cdot 2\pi m ~v(0)\approx\frac{\pi^2}{m}+\frac{\pi^4 \alpha'^2}{2 m^3}+\frac{3 \pi^6 \alpha'^4}{4 m^5}+{\cal O}(\alpha'^6)
\ee
Thus the leading order expansion of $\delta S_T$  precisely gives the exact value calculated in previous section.  However, the higher orders do not match.  This is because that in our approach the backreaction of DBI field shall be weak [1] and perturbation of metric $v(u)$ is samll.  Thus, the result is reliable only in the case of small gravitation entropy, which is the case of leading order approximation.  Note that like as  $S_{DBI}$ the correction $\delta S_H$ is also  divergent as $m\rightarrow 2\pi \alpha'$. 
\section{Conclusions}
In this paper we extend  the calculations of  the generalized gravitational entropy proposed in recent by  Lewkowycz  and Maldacena [1] to the systems of Boson fields, Fermion fields and Maxwell fields which have arbitrary frequency and mode numbers on the BTZ  spacetime.  We find the associated regular wave solution in each case and  use it  to calculate the exact gravitational entropy.  The results show that there is a threshold  frequency below which the Fermion fields  could not contribute the generalized gravitational entropy.  Also, the static and zero-mode solutions  have no entropy,  contrast to that in scalar fields.  We also found that the entropy of  the static scalar fields and non-static fermions  is an increasing function of mode numbers and, after it arrive a maximum entropy it becomes a deceasing function and is derived to the  asymptotic value.  We  calculate the entropy of  the Maxwell fields and use the duality between EM and scalar fields to explain its result.  We also study the back-reaction by D branes, which are described by DBI action, and explicitly check that the classical solution which gives the black hole entropy precisely corrects the black hole area law.

\section{Appendix : Fermion Solution from Conformal Transformation}
We will apply the method in [19,20,21] to find the Fermion solution in BTZ spacetime to check our solution in section 4.  Consider the direct product space 
\be
g_{\mu\nu}dx^\mu dx^\nu=g_{ab}(x)dx^a dx^a+g_{mn}(y)dy^m dy^n
\ee
then the Dirac equation can be seperated by
\be
\gamma^\mu D_\mu\psi=(\gamma^a D_a+\gamma^n D_n)\Psi
\ee 
The relevant quantities under conformal transformations in 3 dimension are
\be
g_{\mu\nu}=\Omega^2 \tilde g_{\mu\nu},~~\Psi=\Omega^{-1}\tilde  \Psi,~~ D_\mu\Psi=\Omega^{-2}\tilde D_\mu\tilde  \Psi
\ee
Therefore the conformal transformations on BTZ spacetime are 
\be
ds^2_3&=&-A(r)^2dt^2+B^2(r)dr^2+r^2d\phi^2=A(r)^2[-dt^2+{B^2\over A^2}dr^2+{r^2\over A^2}d\phi^2]\\
\Psi&=&A^{-1}\tilde  \Psi\\
D_\mu\Psi&=&A^{-2}\tilde D_\mu\tilde  \Psi=A^{-2}(\partial_t+\tilde D_{\mu,2D})\tilde  \Psi
\ee
We can furthermore expressed $\tilde D_{\mu,2D}$ in conformal transformation relation, i.e.
\be
~d\tilde s_{2}^2&=&{B^2\over A^2}dr^2+{r^2\over A^2}d\phi^2={r^2\over A^2}\Big[{B^2\over C^2}dr^2+d\phi^2\Big]\\
\tilde  \Psi&=&\Big({r\over A}\Big)^{-{1\over 2}}\tilde {\tilde \Psi}\\
\tilde D_{\mu,2D}\tilde\Psi&=&\Big({r\over A}\Big)^{-{3\over 2}}~\tilde {\tilde D}_{\mu,2D}\tilde {\tilde \Psi}=\Big({r\over A}\Big)^{-{3\over 2}}~\Big[\tilde {\tilde D}_r+\partial_{\phi}\Big]\tilde {\tilde \Psi}
\ee
Therefore
\be
\Psi&=&{1\over r^{1\over 2}\sqrt A}\tilde {\tilde \Psi}\\
D_\mu\Psi &=&{1\over r^{1\over 2}A\sqrt A}\Big[\partial_t \tilde {\tilde \Psi}+{A\over r}\tilde {\tilde D}_r\tilde {\tilde \Psi}+{\lambda A\over r}\tilde {\tilde \Psi}\Big],~~~with~~\partial_{\phi}\tilde {\tilde \Psi}=\lambda\tilde {\tilde \Psi}
\ee
which implies 
\be
0=\gamma^\mu D_\mu\Psi&=&{1\over C^{1\over 2}A\sqrt A}\Big(\gamma^0\partial_t+ \gamma^r\partial_y+{\lambda A\over r}\Big)\tilde {\tilde \Psi},~~~with~~dy={B\over A}dr
\ee
If the time part solution is $e^{-i\omega t}$ then 
\be
(\partial^2_y-\omega^2)\tilde {\tilde \Psi}&=&\Big({\lambda A \over r}\Big)^2\tilde {\tilde \Psi}
\ee
For the zero mode $\lambda=0$ we can easily find the following solution
\be
y&=&{1\over 2m}\Big(\ln{r-m\over r+m}\Big) \\
\tilde {\tilde \Psi}&=&e^{\pm\omega y}=\Big({r+m\over \sqrt{r^2-m^2}}\Big)^{\pm{\omega \over M}}\\
\Psi&=&{1\over \sqrt r\sqrt{r^2-m^2}}~\Big({r+m\over \sqrt{r^2-m^2}}\Big)^{\pm{\omega \over m}}
\ee
Changing to the coordinate variable  $r =\sqrt {u^2+m^2}$ then
\be
\Psi&=&{1\over\sqrt { u~\sqrt {u^2+m^2}}}\Big({\sqrt {u^2+m^2}+m\over u}\Big)^{\pm{\omega \over m}}
\ee
which is just eq.(4.12) describing the zero-mode solution of fermion fields on BTZ spacetime.
\\
\\
\begin{center} {\bf REFERENCES}\end{center}
\begin{enumerate}
\item  A. Lewkowycz and J. Maldacena, ``Generalized gravitational entropy," JHEP 08 (2013) 090  [arXiv:1304.4926 [hep-th]].
\item  G. W. Gibbons  and S. W. Hawking, ``Action Integrals and Partition Functions in Quantum Gravity'' , Phys. Rev., D15 (1977) 2752.
\item S. N. Solodukhin, ``Entanglement entropy of black holes '', Living Rev. Relativity 14, (2011), 8 [arXiv:1104.3712 [hep-th]].
\item M. Srednicki ``Entropy and area'',  Phys. Rev. Lett., 71 (1993) 666 
[arXiv:hep-th/9303048].
\item  C.G. Callan and F. Wilczek, ``On geometric entropy", Phys. Lett., B333  (1994) 55 [arXiv:hep-th/9401072].
\item  V.P Frolov and I. Novikov, ``Dynamical origin of the entropy of a black hole", Phys. Rev., D48 (1993) 4545 [arXiv:gr-qc/9309001].
\item L.  Susskind, ``Some speculations about black hole entropy in string theory'', (1993) [arXiv:hep-th/9309145 [hep-th]].
\item F. Larsen and F. Wilczek, ``Renormalization of black hole entropy and of the gravitational coupling constant", Nucl. Phys., B458 (1996) 249 [arXiv:hep-th/9506066].
\item S. N. Solodukhin, ``Black hole entropy: statistical mechanics agrees thermodynamics'',  Phys. Rev., D54 (1996) 3900 [arXiv:hep-th/9601154].
\item N. D. Birrell and P. C. W. Davies,``Quantum Fields in Curved Space'', (Cambridge University Press, New York, 1982).
\item  L. Parker and D. Toms ,``Quantum Field Theory in Curved Spacetime'', (Cambridge University Press, New York, 2009). 
\item  M. Banados, C. Teitelboim  and J. Zanelli, `` The Black hole in three dimensional space-time'',  Phys.Rev.Lett., 69 (1992)1849, [arXiv:hep-th/9204099].
\item  M. Banados, M. Henneaux, C. Teitelboim  and J. Zanelli, ``Geometry of the 2+1 Black Hole'', Phys.Rev.D48 (1993) 1506 [arXiv:gr-qc/9302012].
\item  Wung-Hong Huang, ``Generalized Gravitational Entropy of  Interacting Scalar Field and Maxwell Field'' , Physics Letters B. 739 (2014) 365[arXiv:1409.4893 [hep-th]].
\item  D. Tong, ``A Holographic Flat Band'', JHEP11(2011)125 [arXiv:1108.1381[hep-th]].
\item M. Henningson and K. Sfetsos, `` Spinors and the AdS/CFT correspondence'', Phys.Lett. B431 (1998) 63   [arXiv:hep-th/9803251].
\item M.  Henneaux, `` Boundary terms in the AdS/CFT correspondence for spinor fields'', [[arXiv:hep-th/9902137].
\item  C. G. Callan and J. M. Maldacena, ``Brane Dynamics From the Born-Infeld Action", Nucl.Phys. B513 (1998) 198 [arXiv:hep-th/9708147]
\item  G. W.Gibbons, M. Rogatko, ``The Decay of Dirac Hair around a Dilaton Black Hole",  Phys.Rev.D77 (2008)044034  [arXiv:hep-th/0801.3130]
\item  G. W.Gibbons, M. Rogatko,  A. Szyplowska ``Decay of Massive Dirac Hair on a Brane-World Black Hole",   Phys.Rev.D77 (2008)064024  [arXiv:hep-th/0802.3259]
\item R. Moderski, M. Rogatko,``Decay of Dirac Massive Hair in the Background of Spherical Black Hole",   Phys.Rev.D77 (2008)124007  [arXiv:hep-th/0805.0665]
\item R.B. Mann  and S.N. Solodukhin, ``Quantum scalar field on three-dimensional (BTZ) black hole instanton: Heat kernel, effective action and thermodynamics", Phys. Rev. D55 (1997) 3622 [arXiv:hep-th/9609085]. 
\item F. Belgiorno, S. L Cacciatori, F. D. Piazza, O. F Piattella, ``Quantum properties of the Dirac field on BTZ black hole backgrounds",  J. Phys. A44 (2011) 025202 [arXiv:1007.4439  [hep-th]].
\item  D.V. Singh and  S.  Siwach, ``Fermion Fields in BTZ Black Hole Space-Time and Entanglement", Advances in High Energy Physics, 2015 (2015) 528762  [arXiv:1406.3799 [hep-th]]. 
\end{enumerate}
\end{document}